\documentclass[a4paper,final,12pt,journal,onecolumn]{IEEEtran}%
\usepackage{mathrsfs}
\usepackage{amsfonts}
\usepackage{amsmath}
\usepackage{amssymb}
\usepackage{graphicx}%
\providecommand{\U}[1]{\protect\rule{.1in}{.1in}}
\begin{document}

\title{Discrete-Time Output-Feedback Robust Repetitive Control for a Class of
Nonlinear Systems by Additive State Decomposition}
\author{Quan Quan, Lu Jiang and Kai-Yuan Cai \thanks{Corresponding Author: Quan Quan,
Associate Professor, Department of Automatic Control, Beijing University of
Aeronautics and Astronautics, Beijing 100191, qq\_buaa@buaa.edu.cn,
http://quanquan.buaa.edu.cn.}}
\maketitle

\begin{abstract}
The discrete-time robust repetitive control (RC, or repetitive controller,
also designated RC) problem for nonlinear systems is both challenging and
practical. This paper proposes a discrete-time output-feedback RC design for a
class of systems subject to measurable nonlinearities to track reference
robustly with respect to the period variation. The design relies on additive
state decomposition, by which the output-feedback RC problem is decomposed
into an \emph{output-feedback} RC problem for a \emph{linear} time-invariant
system and a \emph{state-feedback} stabilization problem for a
\emph{nonlinear} system. Thanks to the decomposition, existing controller
design methods in both the frequency domain and time domain can be employed to
make the robustness and discretization for a nonlinear system tractable. To
demonstrate the effectiveness, an illustrative example is given.

\end{abstract}


\section{Introduction}

Repetitive Control (RC, or repetitive controller, also designated RC) is a
control method used specifically in tracking or rejecting periodic signals. In
the past two decades, RC for linear time-invariant (LTI) systems has reached
maturity. There has been little research, however, on RC for nonlinear systems
\cite{Quan(2010)}. This is the initial motivation of this paper. One of the
major drawbacks of RC is that the control accuracy is sensitive to period
variation of the external signals. It has been shown in \cite{Steinbuch(2002)}
that, with a period variation as small as $1.5\%$ for an LTI system, the gain
of the internal model part of the RC drops from $\infty$ to $10$. As a result,
the tracking accuracy may be far from satisfactory, especially for
high-precision control. For such a purpose, higher-order RCs composed of
several delay blocks in series were proposed to improve the robustness of the
control accuracy against period variation \cite{Steinbuch(2002)}%
-\cite{Pipeleers(2008)}. However, these methods cannot be applied to nonlinear
systems directly as they are all based on transfer functions and
frequency-domain analysis. This is the second motivation of this paper.
Although controllers are often implemented by digital processors nowadays, it
is rare to see a discrete-time RC design for a continuous nonlinear system.
Unlike LTI systems, nonlinear systems cannot be represented as explicit, exact
discrete-time models. Only approximate controller design methods can be
applied by taking the discretization error as an external disturbance. This
often requires the resulting closed-loop system to be input-to-state stable
(ISS, or input-to-state stability, also designated ISS) with respect to the
discretization error \cite{Nesic(2001)}. However, a linear RC system is a
neutral type system in a critical case \cite{Quan(IET)}. The characteristic
equation of the neutral type system has an infinite sequence of roots with
negative real parts approaching zero. Consequently, the ISS property cannot be
obtained. Therefore, in theory, a discrete-time RC cannot be obtained by
discretizing a continuous RC directly for nonlinear systems. This is the third
motivation. In fact, the discretization of systems in turn brings in
uncertainties in the period. Since the number of delay blocks is an integer,
$T/T_{s}$ has to be rounded to the nearest integer $N$, where $T$ is the
period of the reference signal and $T_{s}$ is the sample time. So, we in fact
take $NT_{s}$ as the known period rather than $T.$ This period variation is
caused by discretization.

Based on the discussion above, a discrete-time robust RC problem for nonlinear
systems is both challenging and practical. In this paper, we will focus on a
discrete-time \emph{output-feedback} robust RC problem\ for\ a class of
systems with measurable nonlinearities. For this problem, we design a
discrete-time output-feedback robust RC under a new tracking framework, named
the \emph{additive-state-decomposition-based tracking control framework
}\cite{Quan(RNC)}. The key idea is to decompose the output-feedback RC problem
into two\emph{ }well-solved\emph{ }control\emph{ }problems by additive state
decomposition\footnote{Additive state decomposition \cite{Quan(RNC)} is
different from the lower-order subsystem decomposition methods existing in the
literature. Concretely, taking the system $\dot{x}\left(  t\right)  =f\left(
t,x\right)  ,x\in%
%TCIMACRO{\U{211d} }%
%BeginExpansion
\mathbb{R}
%EndExpansion
^{n}$ for example, it is decomposed into two subsystems: $\dot{x}_{1}\left(
t\right)  =f_{1}\left(  t,x_{1},x_{2}\right)  $ and $\dot{x}_{2}\left(
t\right)  =f_{2}\left(  t,x_{1},x_{2}\right)  $, where $x_{1}\in%
%TCIMACRO{\U{211d} }%
%BeginExpansion
\mathbb{R}
%EndExpansion
^{n_{1}}\ $and$\ x_{2}\in%
%TCIMACRO{\U{211d} }%
%BeginExpansion
\mathbb{R}
%EndExpansion
^{n_{2}},$ respectively. The lower-order subsystem decomposition satisfies
$n=n_{1}+n_{2}$ and $x=x_{1}\oplus x_{2}.$ By contrast, the proposed additive
state decomposition satisfies $n=n_{1}=n_{2}$ and$\ x=x_{1}+x_{2}.$}: an
output-feedback RC for an LTI system and a state feedback stabilized control
for a nonlinear system. Since the RC problem is only limited to an LTI system,
existing robust higher-order RC methods can be applied directly. Moreover,
according to the properties of the two control problems, we can adopt two
different ways to design discrete-time controllers, i.e., discrete-time model
design for the LTI component and emulation design for the nonlinear component
\cite{Nesic(2001)}. Finally, one can combine the discrete-time output-feedback
robust RC with the discrete-time state-feedback stabilized controller to
achieve the original control goal.

This paper is an extension of our previous paper \cite{Quan(RNC)} and focuses
on the RC problem. Here we propose for a class of nonlinear systems a detailed
RC design. The contributions of this paper are: i) the discrete-time RC
problem is solved for a class of nonlinear systems \emph{for the first time}
(covering the discrete-time output-feedback robust RC problem for a class of
continuous-time nonlinear systems); ii) more importantly, a bridge is built
between existing RC design methods for LTI systems and a class of nonlinear
systems so that more RC problems for nonlinear systems become tractable.

We use the following notation. $%
%TCIMACRO{\U{211d} }%
%BeginExpansion
\mathbb{R}
%EndExpansion
^{n}$ is Euclidean space of dimension $n$. $\left\Vert \mathbf{\cdot
}\right\Vert $ denotes the Euclidean vector norm or induced matrix norm. The
symbol $f\in\mathcal{L}_{\infty}$ implies that $\left\Vert f\right\Vert
_{\infty}\triangleq\sup_{t\in\left[  0,\infty\right)  }\left\Vert f\left(
t\right)  \right\Vert <\infty.$ $\mathcal{L}$ and $\mathcal{L}^{-1}$ denote
Laplace transform and inverse Laplace transform, respectively. $\mathcal{Z}$
and $\mathcal{Z}^{-1}$ denote Z-transform and inverse Z-transform,
respectively. $%
%TCIMACRO{\U{2115} }%
%BeginExpansion
\mathbb{N}
%EndExpansion
$ denotes nonnegative integers. The following definitions can also be found in
\cite{Khalil(2002)}. A continuous function $\alpha:\left[  0,a\right)
\rightarrow\left[  0,\infty\right)  $ is said to belong to class
$\mathcal{K}\ $if$\ $it\ is strictly increasing and $\alpha\left(  0\right)
=0.$ It is said to belong to class $\mathcal{K}_{\infty}\ $if $a=\infty$ and
$\alpha\left(  r\right)  \rightarrow\infty$ as $r\rightarrow\infty.$ A
continuous function $\beta:\left[  0,a\right)  \times\left[  0,\infty\right)
\rightarrow\left[  0,\infty\right)  $ is said to belong to class
$\mathcal{KL}$ if, for each fixed $s,$ the mapping $\beta\left(  r,s\right)  $
belongs to $\mathcal{K}$ with respect to $r$ and, for each fixed $r$, the
mapping $\beta\left(  r,s\right)  $ is decreasing with respect with $s$ and
$\beta\left(  r,s\right)  \rightarrow0$ as $s\rightarrow\infty$.

\section{Problem Formulation}

Consider a class of single-input-single-output (SISO) nonlinear systems
\cite{Marino(1995)}:%
\begin{align}
\dot{x}  &  =Ax+bu+\phi\left(  y\right)  +d,x\left(  0\right)  =x_{0}%
\nonumber\\
y  &  =c^{T}x \label{equ0}%
\end{align}
where $A\in%
%TCIMACRO{\U{211d} }%
%BeginExpansion
\mathbb{R}
%EndExpansion
^{n\times n}\ $is a known stable constant matrix (see \textit{Remark 1}),
$b\in%
%TCIMACRO{\U{211d} }%
%BeginExpansion
\mathbb{R}
%EndExpansion
^{n}\ $and $c\in%
%TCIMACRO{\U{211d} }%
%BeginExpansion
\mathbb{R}
%EndExpansion
^{n}\ $are known constant vectors, $\phi:%
%TCIMACRO{\U{211d} }%
%BeginExpansion
\mathbb{R}
%EndExpansion
\rightarrow%
%TCIMACRO{\U{211d} }%
%BeginExpansion
\mathbb{R}
%EndExpansion
^{n}$ is a known nonlinear function vector, $x\left(  t\right)  \in%
%TCIMACRO{\U{211d} }%
%BeginExpansion
\mathbb{R}
%EndExpansion
^{n}$ is the state vector, $y\left(  t\right)  \in%
%TCIMACRO{\U{211d} }%
%BeginExpansion
\mathbb{R}
%EndExpansion
$ is the output, $u\in%
%TCIMACRO{\U{211d} }%
%BeginExpansion
\mathbb{R}
%EndExpansion
$ is the control, and $d\in%
%TCIMACRO{\U{211d} }%
%BeginExpansion
\mathbb{R}
%EndExpansion
^{n}$ is an unknown periodic bounded signal with period $T>0$. The reference
$r\left(  t\right)  \in%
%TCIMACRO{\U{211d} }%
%BeginExpansion
\mathbb{R}
%EndExpansion
$ is known and sufficiently smooth with period $T$. It is assumed that only
$y\left(  t\right)  $ is available from measurements. In this paper, we
consider the continuous-time system (\ref{equ0}) by using a discrete-time
controller with a sampling period $T_{s}>0,$ where $T=NT_{s},$ $N\in%
%TCIMACRO{\U{2115} }%
%BeginExpansion
\mathbb{N}
%EndExpansion
.$ More precisely, $u$ in (\ref{equ0}) is constant during a sampling interval,
so that $u\left(  t\right)  =u\left(  kT_{s}\right)  =:$ $u\left(  k\right)
,$ $t\in\left[  kT_{s},\left(  k+1\right)  T_{s}\right)  ,$ $k\in%
%TCIMACRO{\U{2115} }%
%BeginExpansion
\mathbb{N}
%EndExpansion
$. In practice, $T$ is not known exactly or is varying, namely period $T$ is
uncertain. On the other hand, we take $NT_{s}$ instead of $T$ as the period in
the discrete-time controller design. Since$\ NT_{s}\neq T$ in general, $T$ can
be also considered as a variation of $NT_{s}$.

\textbf{Assumption 1}. The pair $\left(  A,c\right)  $ is observable.

Under \textit{Assumption 1},\textit{\ }the objective is to design a
discrete-time output-feedback RC for the nonlinear system (\ref{equ0}) such
that $y-r$ is uniformly ultimately bounded with the ultimate bound being
robust with respect to the period variation \footnote{Let $T=NT_{s}+\Delta$ be
the true period, where $\Delta>0$ is the perturbation. By using $NT_{s}\ $in
the design, $y-r$ is uniformly ultimately bounded with the ultimate bound
$d_{e_{\Delta}}>0.$ Here robustness can be roughly understood to mean that
$\frac{d_{e_{\Delta}}}{\left\vert \Delta\right\vert }$ is small.}.

\textbf{Remark 1}\textit{.} Under \textit{Assumption 1}, there always exists a
vector $p\in%
%TCIMACRO{\U{211d} }%
%BeginExpansion
\mathbb{R}
%EndExpansion
^{n}$ such that $A+pc^{T}$ is stable, whose eigenvalues can be assigned
freely. As a result, (\ref{equ0}) can be rewritten as $\dot{x}=\left(
A+pc^{T}\right)  x+bu+\left(  \phi\left(  y\right)  -py\right)  +d.$
Therefore, without loss of generality, we assume $A$ to be stable.

\textbf{Remark 2}\textit{.} The nonlinear function vector $\phi\ $can be
arbitrary. Here, we do not specify its form. Moreover, the nonlinear system
(\ref{equ0}) is allowed to be a non-minimum phase\textit{\ }system
\cite{Quan(RNC)}.

Before proceeding further, the following preliminary result on ISS\textbf{ }is required.

\textbf{Definition 1 }\cite{Nesic(2001)}. The system
\begin{equation}
\dot{x}=f\left(  x,u\left(  x\right)  ,d_{c}\right)  \label{iss_sys}%
\end{equation}
is ISS with respect to $d_{c}$\ if there exist $\beta\in\mathcal{KL}%
\ $and$\ \gamma\in\mathcal{K}$ such that the solutions of the system satisfy
$\left\Vert x\left(  t\right)  \right\Vert $ $\leq$ $\beta\left(  \left\Vert
x\left(  0\right)  \right\Vert ,t\right)  $ $+\gamma\left(  \left\Vert
d_{c}\right\Vert _{\infty}\right)  $, $\forall x\left(  0\right)  ,$ $d_{c}%
\in\mathcal{L}_{\infty},$ $\forall t\geq0$.

Suppose that the feedback is implemented by a sampler and zero-order hold as%
\begin{equation}
u\left(  t\right)  =u\left(  x\left(  k\right)  \right)  ,t\in\left[
kT_{s},\left(  k+1\right)  T_{s}\right)  ,k\in%
%TCIMACRO{\U{2115} }%
%BeginExpansion
\mathbb{N}
%EndExpansion
. \label{iss_u_dis}%
\end{equation}
Then, we have

\textbf{Theorem 1 }\cite{Nesic(2001)}. If the continuous-time system
(\ref{iss_sys}) is ISS, then there exist $\beta\in\mathcal{KL}\ $%
and\ $\gamma\in\mathcal{K}$ such that given any triple of strictly positive
numbers $\left(  \Delta_{x},\Delta_{d_{c}},\nu\right)  ,$ there exists
$T^{\ast}>0$ such that for all $T_{s}\in\left(  0,T^{\ast}\right)  ,$
$\left\Vert x\left(  0\right)  \right\Vert \leq\Delta_{x},$ $\left\Vert
d_{c}\right\Vert _{\infty}\leq\Delta_{d_{c}},$ the solutions of the
sampled-data system $\dot{x}=f\left(  x,u\left(  x\left(  k\right)  \right)
,d_{c}\right)  $ satisfy:%
\begin{equation}
\left\Vert x\left(  k\right)  \right\Vert \leq\beta\left(  \left\Vert x\left(
0\right)  \right\Vert ,kT_{s}\right)  +\gamma\left(  \left\Vert d_{c}%
\right\Vert _{\infty}\right)  +\nu,k\in%
%TCIMACRO{\U{2115} }%
%BeginExpansion
\mathbb{N}
%EndExpansion
. \label{semiISS}%
\end{equation}

\textbf{Remark 3}\textit{. Theorem 1} states that if the continuous-time
closed-loop system is ISS, then the sampled-data system with the emulated
controller will be semiglobally practically ISS with a sufficiently small
$T_{s}$.

\section{Discrete-Time Output-Feedback Robust RC by Additive State
Decomposition}

\subsection{Additive State Decomposition}

In order to make the paper self-contained, the additive state decomposition of
(\ref{equ0}) in \cite{Quan(RNC)} is recalled here briefly. Consider the system
(\ref{equ0}) as the original system. We choose the primary system as follows:%
\begin{align}
\dot{x}_{p}  &  =Ax_{p}+bu_{p}+\phi\left(  r\right)  +d\nonumber\\
y_{p}  &  =c^{T}x_{p},x_{p}\left(  0\right)  =x_{0}. \label{equ1_Pri}%
\end{align}
Then the secondary system is determined by the original system (\ref{equ0})
and the primary system (\ref{equ1_Pri}) as%
\begin{align}
\dot{x}_{s}  &  =Ax_{s}+bu_{s}+\phi\left(  y\right)  -\phi\left(  r\right)
\nonumber\\
y_{s}  &  =c^{T}x_{s},x_{s}\left(  0\right)  =0 \label{equ1_Sec}%
\end{align}
where $u_{s}=u-u_{p}.$ According to the additive state decomposition, we have%
\begin{equation}
x=x_{p}+x_{s}\text{ and }y=y_{p}+y_{s}. \label{equ1_relation}%
\end{equation}
The secondary system (\ref{equ1_Sec}) is further written as%
\begin{align}
\dot{x}_{s}  &  =Ax_{s}+bu_{s}+\phi\left(  r+y_{s}+e_{p}\right)  -\phi\left(
r\right) \nonumber\\
y_{s}  &  =c^{T}x_{s},x_{s}\left(  0\right)  =0 \label{equ1_Sec1}%
\end{align}
where $e_{p}\triangleq y_{p}-r.$ If $e_{p}\equiv0,$ then $\left(  x_{s}%
,u_{s}\right)  =0$ is an equilibrium point of (\ref{equ1_Sec1}).

Controller design for the decomposed systems (\ref{equ1_Pri}) and
(\ref{equ1_Sec}) will use their outputs or states as feedback. However, they
are unknown. For such a purpose, an observer is proposed to estimate $y_{p}$
and $x_{s}.$

\textbf{\textbf{Theorem 2 }}\cite{Quan(RNC)}.\textit{ }Suppose that an
observer is designed to estimate $y_{p}$ and $x_{s}$ in (\ref{equ1_Pri}%
)-(\ref{equ1_Sec}) as follows:%
\begin{align}
\hat{y}_{p}  &  =y-c^{T}\hat{x}_{s}\label{equ1_Obs1}\\
\dot{\hat{x}}_{s}  &  =A\hat{x}_{s}+bu_{s}+\phi\left(  y\right)  -\phi\left(
r\right)  ,\hat{x}_{s}\left(  0\right)  =0. \label{equ1_Obs2}%
\end{align}
Then $\hat{y}_{p}\equiv y_{p}$ and $\hat{x}_{s}\equiv x_{s}.$

\textbf{Remark 4}. Additive state decomposition brings in two benefits. First,
since output of the primary system and state of the secondary system can be
observed, the original tracking problem for the system (\ref{equ0}) is
correspondingly decomposed into two problems: an \emph{output-feedback}
tracking problem for an LTI `primary' system ($y_{p}\rightarrow r$) and a
\emph{state-feedback} stabilization problem for the complementary `secondary'
system ($x_{s}\rightarrow0$). As a result, we have $y\rightarrow r$ according
to (\ref{equ1_relation}). Since the tracking task is only assigned to the LTI
system, it is therefore much easier than that for the nonlinear system
(\ref{equ0}). The state-feedback stabilization is also easier than the
output-feedback stabilization as the non-minimum phase problem is avoided.
Secondly, for the two decomposed components, different discrete-time
controller design methods can be employed (shown in Fig. 1) because for LTI
systems an explicit, exact discrete-time model can be obtained whereas for
nonlinear systems it cannot. The ISS property cannot be obtained for a
traditional RC system. This implies that a sufficiently small uncertainty may
cause instability. So, it is appropriate to follow the discrete-time model
design for the discrete-time RC design of the linear primary system. On the
other hand, a state-feedback stabilization problem for the secondary system is
independent of the internal model of RCs. The resultant closed-loop system can
be rendered ISS. So, we can adopt the emulation design for the discrete-time
controller design of the nonlinear secondary system.\begin{figure}[h]
\begin{center}
\includegraphics[
scale=0.8 ]{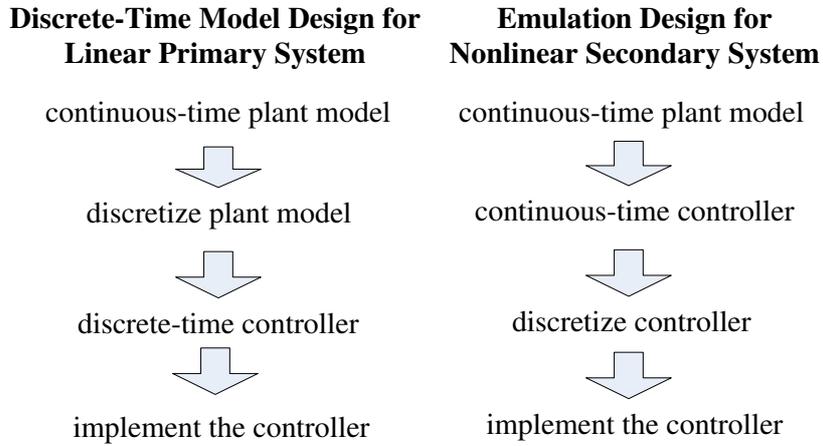}
\end{center}
\par
\vspace*{-5pt}\caption{Different discrete-time controller design for the
primary system (\ref{equ1_Pri}) and the secondary system (\ref{equ1_Sec1})}%
\end{figure}\vspace*{-18pt}

\subsection{Controller Design for Primary System and Secondary System}

So far, we have decomposed the system into two systems each with its own task.
In this section, we investigate controller design for each.

\textbf{Problem 1}. \textit{For (\ref{equ1_Pri}), design a discrete-time
output-feedback RC as}%
\begin{equation}
u_{p}\left(  k\right)  =\mathcal{Z}^{-1}\left(  G\left(  z\right)
e_{p}\left(  z\right)  \right)  \label{repetitivecontroller0}%
\end{equation}
\textit{such that }$e_{p}\left(  k\right)  =r\left(  k\right)  -y_{p}\left(
k\right)  \rightarrow\mathcal{B}\left(  \delta\right)  $
\footnote{$\mathcal{B}\left(  \delta\right)  \triangleq\left\{  \xi\in%
%TCIMACRO{\U{211d} }%
%BeginExpansion
\mathbb{R}
%EndExpansion
\left\vert \left\Vert \xi\right\Vert \leq\delta\right.  \right\}  ,$
$\delta>0;$ the notation $x\left(  k\right)  \rightarrow\mathcal{B}\left(
\delta\right)  $ means $\underset{y\in\mathcal{B}\left(  \delta\right)  }%
{\min}$ $\left\vert x\left(  k\right)  -y\right\vert \rightarrow0\ $as
$k\rightarrow\infty.$}\textit{\ as }$k\rightarrow\infty,$ \textit{where
}$\delta=\delta\left(  r,d\right)  >0$\textit{ depends on the reference }%
$r$\textit{ and disturbance }$d$\textit{.}

Since (\ref{equ1_Pri}) is an LTI system, it can be written as%
\begin{equation}
y_{p}\left(  s\right)  =P\left(  s\right)  u_{p}\left(  s\right)
+d_{r}\left(  s\right)  \label{continuous}%
\end{equation}
where $P\left(  s\right)  =c^{T}\left(  sI-A\right)  ^{-1}b$ and$\ d_{r}%
\left(  s\right)  =c^{T}\left(  sI-A\right)  ^{-1}\mathcal{L}\left(
\phi\left(  r\left(  t\right)  \right)  +d\left(  t\right)  \right)  .$ Then
by using zero-order hold on the input with the sampling period $T_{s}$, the
continuous-time LTI system (\ref{continuous}) is discretized exactly as%
\begin{equation}
y_{p}\left(  z\right)  =P\left(  z\right)  u_{p}\left(  z\right)
+d_{r}\left(  z\right)  \label{discrete}%
\end{equation}
where $P\left(  z\right)  =c^{T}\left(  zI-F\right)  ^{-1}Hb,$ $F=e^{AT_{s}%
},H=\int_{0}^{T_{s}}e^{As}ds.$ Similar to \cite{Steinbuch(2007)}%
,\cite{Pipeleers(2008)}, we design a discrete-time output-feedback RC in the
form of (\ref{repetitivecontroller0}) as%
\begin{equation}
u_{p}\left(  z\right)  =\left(  1+L\left(  z\right)  \frac{Q\left(  z\right)
W\left(  z\right)  z^{-N}}{1-Q\left(  z\right)  W\left(  z\right)  z^{-N}%
}\right)  e_{p}\left(  z\right)  \label{repetitivecontroller}%
\end{equation}
where $W\left(  z\right)  $ is the gain adjusting or the higher-order RC
function, given by%
\begin{equation}
W\left(  z\right)  =\underset{i=1}{\overset{p}{%
%TCIMACRO{\dsum }%
%BeginExpansion
{\displaystyle\sum}
%EndExpansion
}}w_{i}z^{-\left(  i-1\right)  N} \label{W}%
\end{equation}
with $\underset{i=1}{\overset{p}{%
%TCIMACRO{\dsum }%
%BeginExpansion
{\displaystyle\sum}
%EndExpansion
}}w_{i}=1.$ For a traditional RC, $W\left(  z\right)  =1.$ The stability of
the closed-loop system corresponding to (\ref{discrete}) and
(\ref{repetitivecontroller}) is given by \textit{Theorem 3.}

\textbf{Theorem 3}. Let $u_{p}$ in (\ref{discrete}) be designed as in
(\ref{repetitivecontroller}). Suppose i) $\frac{1}{1+P\left(  z\right)
},P\left(  z\right)  ,L\left(  z\right)  ,Q\left(  z\right)  $ are stable, ii)%
\begin{equation}
\left\vert Q\left(  z\right)  W\left(  z\right)  z^{-N}\left(  1-T\left(
z\right)  L\left(  z\right)  \right)  \right\vert <1,\forall\left\vert
z\right\vert =1 \label{condition0}%
\end{equation}
where $T\left(  z\right)  =\frac{P\left(  z\right)  }{1+P\left(  z\right)  }.$
Then the tracking error $e_{p}$ is uniformly ultimately bounded. Furthermore,
if $\mathcal{Z}^{-1}\left(  \left(  1-Q\left(  z\right)  z^{-N}\right)
\left(  r\left(  z\right)  -d_{r}\left(  z\right)  \right)  \right)
\rightarrow0,$ then $e_{p}\left(  k\right)  =r\left(  k\right)  -y_{p}\left(
k\right)  \rightarrow0$ \textit{as }$k\rightarrow\infty.$

\textit{Proof}. By substituting (\ref{repetitivecontroller}) into
(\ref{discrete}), the tracking error of the primary system can be written as%
\begin{equation}
e_{p}\left(  z\right)  =\frac{1}{1+P\left(  z\right)  }\frac{\left(
1-Q\left(  z\right)  W\left(  z\right)  z^{-N}\right)  \left(  r\left(
z\right)  -d_{r}\left(  z\right)  \right)  }{1-Q\left(  z\right)  W\left(
z\right)  z^{-N}\left(  1-T\left(  z\right)  L\left(  z\right)  \right)  }
\label{trackingerror}%
\end{equation}
where $T\left(  z\right)  =\frac{P\left(  z\right)  }{1+P\left(  z\right)  }.$
A sufficient criterion for stability of the closed-loop system now becomes
that $\frac{1}{1+P\left(  z\right)  }$ and $\frac{1-Q\left(  z\right)
W\left(  z\right)  z^{-N}}{1-Q\left(  z\right)  W\left(  z\right)
z^{-N}\left(  1-T\left(  z\right)  L\left(  z\right)  \right)  }$ are both
stable. The transfer function $\frac{1}{1+P\left(  z\right)  }$ is stable by
condition i). For stability of $\frac{1-Q\left(  z\right)  W\left(  z\right)
z^{-N}}{1-Q\left(  z\right)  W\left(  z\right)  z^{-N}\left(  1-T\left(
z\right)  L\left(  z\right)  \right)  }$, to apply the small gain theorem,
$Q\left(  z\right)  W\left(  z\right)  z^{-N}\left(  1-T\left(  z\right)
L\left(  z\right)  \right)  \ $is required to be stable first. This requires
that $\frac{1}{1+P\left(  z\right)  },$ $P\left(  z\right)  \ $and$\ Q\left(
z\right)  $ are stable, which are satisfied by given conditions. Therefore, if
(\ref{condition0}) holds, then $\frac{1-Q\left(  z\right)  W\left(  z\right)
z^{-N}}{1-Q\left(  z\right)  W\left(  z\right)  z^{-N}\left(  1-T\left(
z\right)  L\left(  z\right)  \right)  }$ is stable by the small gain theorem.
Then the tracking error $e_{p}$ is uniformly ultimately bounded. Furthermore,
taking $\left(  1-Q\left(  z\right)  W\left(  z\right)  z^{-N}\right)  \left(
r\left(  z\right)  -d_{r}\left(  z\right)  \right)  $ as a new input, since
$\frac{1}{1+P\left(  z\right)  }\frac{1}{1-Q\left(  z\right)  W\left(
z\right)  z^{-N}\left(  1-T\left(  z\right)  L\left(  z\right)  \right)  }$ is
stable, $e_{p}\left(  k\right)  =r\left(  k\right)  -y_{p}\left(  k\right)
\rightarrow0$ if
\[
\mathcal{Z}^{-1}\left(  1-Q\left(  z\right)  z^{-N}\left(  r\left(  z\right)
-d_{r}\left(  z\right)  \right)  \right)  \rightarrow0
\]
\textit{as }$k\rightarrow\infty.\ \square$

From \textit{Theorem 3}, one can see that the stability depends on three main
elements of the controller (\ref{repetitivecontroller}): $L\left(  z\right)
,Q\left(  z\right)  \ $and$\ W\left(  z\right)  $. The ideal design is to let%
\begin{equation}
1-T\left(  z\right)  L\left(  z\right)  =0,Q\left(  z\right)  =1.
\label{condition}%
\end{equation}
As a result, the condition (\ref{condition0}) is satisfied and
(\ref{trackingerror}) becomes%
\[
e_{p}\left(  z\right)  =\frac{1}{1+P\left(  z\right)  }\left(  1-W\left(
z\right)  z^{-N}\right)  \left(  r\left(  z\right)  -d_{r}\left(  z\right)
\right)  .
\]
As a result, $e_{p}\left(  k\right)  =r\left(  k\right)  -y_{p}\left(
k\right)  \rightarrow0$ with $\mathcal{Z}^{-1}\left(  \left(  1-W\left(
z\right)  z^{-N}\right)  \left(  r\left(  z\right)  -d_{r}\left(  z\right)
\right)  \right)  \rightarrow0$\ as $k\rightarrow\infty.$ However,
(\ref{condition}) often cannot be satisfied. In the following, we will discuss
how to design $L\left(  z\right)  ,Q\left(  z\right)  \ $and$\ W\left(
z\right)  $ in practice.

\textbf{Remark 5 (on design of }$L\left(  z\right)  ).$ In practice, the
transfer function $T\left(  z\right)  $ might be non-minimum phase, or its
relative degree is nonzero. So, we cannot find an $L\left(  z\right)  $ to
satisfy $T\left(  z\right)  L\left(  z\right)  =1$ exactly. Alternatively, the
filter $L\left(  z\right)  $ can be designed by using the zero phase error
tracking controller (ZPETC) algorithm as proposed in \cite{Tomizuka(1987)}.
When there are zeros outside the unit circle for $T\left(  z\right)  $, one
can rewrite $T\left(  z\right)  $ in the following form%
\[
T\left(  z\right)  =\frac{z^{-n_{T}}T_{n}\left(  z^{-1}\right)  }{T_{d}\left(
z^{-1}\right)  }=\frac{z^{-n_{T}}T_{n}^{+}\left(  z^{-1}\right)  T_{n}%
^{-}\left(  z^{-1}\right)  }{T_{d}\left(  z^{-1}\right)  }%
\]
where $T_{n}^{+}\left(  z^{-1}\right)  $ is the cancelable part containing
only the stable zeros, $T_{n}^{-}\left(  z^{-1}\right)  $ is the uncancelable
part containing only the unstable zeros, and $n_{T}$ is the difference between
the order of the numerator and that of the denominator. Based on the
decomposition, the filter $L\left(  z\right)  $ is designed as%
\begin{equation}
L\left(  z\right)  =\frac{T_{d}\left(  z^{-1}\right)  T_{n}^{-}\left(
z\right)  }{\left(  T_{n}^{-}\left(  1\right)  \right)  ^{2}T_{n}^{+}\left(
z^{-1}\right)  } \label{ZPETC_L}%
\end{equation}
where $T_{n}^{-}\left(  z\right)  $ is obtained by replacing each $z^{-1}$ in
$T_{n}^{-}\left(  z^{-1}\right)  $ by $z$. Such a design assures that the
phase of $T\left(  e^{i\omega T_{s}}\right)  L\left(  e^{i\omega T_{s}%
}\right)  $ is $0$ for all frequencies $\omega$, and the gain is $1$ for low frequencies.

\textbf{Remark 6 (on design of }$Q\left(  z\right)  )$. The design of
$L\left(  z\right)  $ has assured that $T\left(  z\right)  L\left(  z\right)
\approx1$ in the low frequency band so that the stability criterion
(\ref{condition0}) holds in the low frequency band. However, the stability
criterion may be violated in the high frequency band. Based on the choice of
$L\left(  z\right)  ,$ the filter $Q\left(  z\right)  $ is chosen to be a
low-pass filter which aims to attenuate the term $\left\vert Q\left(
z\right)  W\left(  z\right)  z^{-N}\left(  1-T\left(  z\right)  L\left(
z\right)  \right)  \right\vert $ in the high frequency band. On the other
hand, by (\ref{trackingerror}), the term $1-Q\left(  z\right)  W\left(
z\right)  z^{-N}$ will determine the tracking performance directly. It is well
known that the low frequency band is often dominant in the signal $r\left(
z\right)  -d_{r}\left(  z\right)  .$ So, an appropriate filter $Q\left(
z\right)  \ $will make $\left\vert \left(  1-Q\left(  z\right)  W\left(
z\right)  z^{-N}\right)  \left(  r\left(  z\right)  -d_{r}\left(  z\right)
\right)  \right\vert \approx0.$ In practice, the trade-off between the
stability and tracking performance must be taken into consideration to seek a balance.

\textbf{Remark 7 (on design of }$W\left(  z\right)  )$. Gains at the harmonics
are expected to be infinite \cite{Steinbuch(2002)}, so $%
%TCIMACRO{\dsum \nolimits_{i=1}^{p}}%
%BeginExpansion
{\displaystyle\sum\nolimits_{i=1}^{p}}
%EndExpansion
w_{i}=1.$ With the redundant freedom, we can design appropriate weighting
coefficients $w_{1},w_{2},...w_{p}\ $to improve the robustness of the tracking
accuracy with respect to the period variation of $r-d_{r}.$

So far, we have designed a discrete-time output-feedback robust RC for
\textit{Problem 1}. In the following we consider the design of a discrete-time
controller for the nonlinear system (\ref{equ1_Sec1}).

\textbf{Problem 2}. \textit{For (\ref{equ1_Sec1}), design a controller }%
\begin{equation}
u_{s}(k)=u_{s}(x_{s}(k)) \label{con_second_dis}%
\end{equation}
\textit{ such that the closed loop system is ISS with respect to the input
}$e_{p}(k)$\textit{, namely}%
\begin{equation}
\left\Vert x_{s}\left(  k\right)  \right\Vert \leq\gamma\left(  \left\Vert
e_{p}\left(  k\right)  \right\Vert \right)  +\nu,k\in%
%TCIMACRO{\U{2115} }%
%BeginExpansion
\mathbb{N}
%EndExpansion
\label{iss_second}%
\end{equation}
\textit{where }$\gamma$\textit{ is a class }$\mathcal{K}$\textit{ function and
}$\nu>0$\textit{ can be made small by reducing the sampling period }$T_{s}%
$\textit{.}

For the secondary system (\ref{equ1_Sec1}), we design a locally Lipschitz
static state feedback%
\begin{equation}
u_{s}(t)=u_{s}(x_{s}(t)). \label{con_second}%
\end{equation}
Then substituting it into (\ref{equ1_Sec1}) yields%
\begin{equation}
\dot{x}_{s}=f\left(  t,x_{s},e_{p}\right)  \label{equ1_Sec1_cld}%
\end{equation}
where $f\left(  t,x_{s},e_{p}\right)  =Ax_{s}+bu_{s}(x_{s}(t))+\phi\left(
r\left(  t\right)  +c^{T}x_{s}+e_{p}\left(  t\right)  \right)  -\phi\left(
r\left(  t\right)  \right)  .$ With respect to the ISS problem for
(\ref{equ1_Sec1_cld}), we have the following theorem.

\textbf{Theorem 4}. Suppose that there exists a continuously differentiable
function $V:\left[  0,\infty\right)  \times%
%TCIMACRO{\U{211d} }%
%BeginExpansion
\mathbb{R}
%EndExpansion
^{n}\rightarrow%
%TCIMACRO{\U{211d} }%
%BeginExpansion
\mathbb{R}
%EndExpansion
$ such that
\begin{align*}
\alpha_{1}\left(  \left\Vert x_{s}\right\Vert \right)   &  \leq V\left(
t,x_{s}\right)  \leq\alpha_{2}\left(  \left\Vert x_{s}\right\Vert \right) \\
\frac{\partial V}{\partial t}+\frac{\partial V}{\partial x_{s}}f\left(
t,x_{s},e_{p}\right)   &  \leq-W\left(  x_{s}\right)  ,\forall\left\Vert
x_{s}\right\Vert \geq\rho\left(  \left\Vert e_{p}\right\Vert \right)  >0
\end{align*}
$\forall\left(  t,x_{s},e_{p}\right)  \in\left[  0,\infty\right)  \times%
%TCIMACRO{\U{211d} }%
%BeginExpansion
\mathbb{R}
%EndExpansion
^{n}\times%
%TCIMACRO{\U{211d} }%
%BeginExpansion
\mathbb{R}
%EndExpansion
,$ where $\alpha_{1},\alpha_{2}$ are class $\mathcal{K}_{\mathcal{\infty}}$
functions, $\rho$ is a class $\mathcal{K}$ function, and $W\left(  x\right)  $
is a continuous positive definite function on $%
%TCIMACRO{\U{211d} }%
%BeginExpansion
\mathbb{R}
%EndExpansion
^{n}.$ Then, given any triple of strictly positive numbers $\left(  \Delta
_{x},\Delta_{d_{c}},\nu\right)  ,$ there exists $T^{\ast}>0$ such that for all
$T_{s}\in\left(  0,T^{\ast}\right)  ,$ $\left\Vert x_{s}\left(  0\right)
\right\Vert \leq\Delta_{x},$ $\left\Vert e_{p}\right\Vert _{\infty}\leq
\Delta_{d_{c}},$ the solutions of the sampled-data system (\ref{equ1_Sec1}),
(\ref{con_second_dis}) satisfy (\ref{iss_second}), where $\gamma=\alpha
_{1}^{-1}\circ\alpha_{2}\circ\rho.$

\textit{Proof.} We can imitate the proof of \textit{Theorem 4.19} in \cite[p.
176]{Khalil(2002)} to show that the continuous-time system
(\ref{equ1_Sec1_cld}) is ISS with $\gamma=\alpha_{1}^{-1}\circ\alpha_{2}%
\circ\rho$. Then, based on \textit{Theorem 1}, we can conclude that the
solutions of the sampled-data system (\ref{equ1_Sec1}), (\ref{con_second_dis})
are semiglobally practically ISS like (\ref{semiISS}). Notice that the term
$\beta\left(  \left\Vert x_{s}\left(  0\right)  \right\Vert ,kT_{s}\right)  $
does not appear in (\ref{iss_second}), as $x_{s}\left(  0\right)  =0\ $implies
$\beta\left(  \left\Vert x_{s}\left(  0\right)  \right\Vert ,kT_{s}\right)
\equiv0$, where $\beta\in\mathcal{KL}.$ $\square$

\textbf{Remark 8}. By \textit{Theorem 4}, if the continuous-time closed loop
system is ISS by (\ref{con_second}), then there exists a sampling period
$T_{s}>0$ such that the resulting closed-loop system by (\ref{con_second_dis})
is semiglobally practically ISS. It is difficult to give an exact $T^{\ast}$
to ensure that \textit{Theorem 4 }is\textit{ }satisfied. Even if $T^{\ast}$ is
given for a general case, it will be conservative. In practice, the sampling
period can be determined by simulation and experiment case-by-case.

\subsection{Controller Integration}

With the two designed controllers (\ref{repetitivecontroller}) and
(\ref{con_second_dis}) for the two subsystems, we can combine them together to
solve the original problem. The result is stated in \textit{Theorem 5}.

\textbf{Theorem 5. }Suppose\textbf{\ }i) \textit{Problems 1-2} are solved; ii)
the observer-controller for system (\ref{equ0}) is designed as:%
\begin{equation}%
\begin{array}
[c]{lll}%
\hat{y}_{p}\left(  k\right)  & = & y\left(  k\right)  -c^{T}\hat{x}_{s}\left(
k\right) \\
\hat{x}_{s}\left(  k+1\right)  & = & F\hat{x}_{s}\left(  k\right)
+Hbu_{s}\left(  k\right)  +H\int_{0}^{T_{s}}e^{As}\left(  y\left(  s\right)
-r\left(  s\right)  \right)  ds,\hat{x}_{s}\left(  0\right)  =0
\end{array}
\label{Pb3Observer}%
\end{equation}
and%
\begin{align}
u_{p}\left(  k\right)   &  =\mathcal{Z}^{-1}\left(  \left(  1+L\left(
z\right)  \frac{Q\left(  z\right)  W\left(  z\right)  z^{-N}}{1-Q\left(
z\right)  W\left(  z\right)  z^{-N}}\right)  \left(  r\left(  z\right)
-\hat{y}_{p}\left(  z\right)  \right)  \right) \nonumber\\
u_{s}\left(  k\right)   &  =u_{s}(\hat{x}_{s}\left(  k\right)
)\label{Pb3Controller}\\
u\left(  k\right)   &  =u_{p}\left(  k\right)  +u_{s}\left(  k\right)
.\nonumber
\end{align}
Then the output of system (\ref{equ0}) satisfies that $y\left(  k\right)
-r\left(  k\right)  \rightarrow\mathcal{B(}\delta+\left\Vert c\right\Vert
\gamma(\delta)+\left\Vert c\right\Vert \nu)$ as $k\rightarrow\infty$.

\textit{Proof. }By \textit{Theorem 2}, the estimates in the observer
(\ref{Pb3Observer}) satisfy $\hat{x}_{p}\equiv x_{p}$ and $\hat{x}_{s}\equiv
x_{s}$. Then the controller $u_{p}$ in (\ref{Pb3Controller}) can drive
$e_{p}\left(  k\right)  =y_{p}\left(  k\right)  -r\left(  k\right)
\rightarrow\mathcal{B(}\delta)$ as $k\rightarrow\infty$ thanks to
\textit{Problem 1 }being solved. In the following, we will further show that
the controller $u_{s}$ in (\ref{Pb3Controller}) can drive $y_{s}\left(
k\right)  \rightarrow\mathcal{B(}\left\Vert c\right\Vert \gamma(\delta
)+\left\Vert c\right\Vert \nu)$ as $k\rightarrow\infty.$ We can conclude this
proof$\ $as $u=u_{p}+u_{s}$ and $y=y_{p}+y_{s}.$ \textit{Proof of }%
$y_{s}\left(  k\right)  \rightarrow\mathcal{B}(\left\Vert c\right\Vert
\gamma(\delta)+\left\Vert c\right\Vert \nu).$ Suppose \textit{Problem 2} is
solved. According to (\ref{iss_second}), we have $\left\Vert y_{s}%
(k)\right\Vert \leq\left\Vert c\right\Vert \left\Vert x_{s}(k)\right\Vert
\leq\left\Vert c\right\Vert \gamma(\left\Vert e_{p}(k)\right\Vert )+\left\Vert
c\right\Vert \nu,k\in%
%TCIMACRO{\U{2115} }%
%BeginExpansion
\mathbb{N}
%EndExpansion
.$ Based on the result of i), we get $e_{p}\rightarrow\mathcal{B(}\delta)$ as
$k\rightarrow\infty.$ This implies that $\left\Vert e_{p}\left(  k\right)
\right\Vert \leq\delta+\varepsilon$ when $k\geqslant N_{0}.$ Then $\left\Vert
y_{s}(k)\right\Vert \leq\left\Vert c\right\Vert \beta(\left\Vert x_{s}%
(N_{0})\right\Vert ,k-N_{0})+\left\Vert c\right\Vert \gamma(\delta
+\varepsilon)+\left\Vert c\right\Vert \nu,k\geqslant N_{0}.$ Since $\left\Vert
c\right\Vert \beta(\left\Vert x_{s}(N_{0})\right\Vert ,k-N_{0})\rightarrow0$
as $k\rightarrow\infty$ and $\varepsilon$ can be chosen arbitrarily small, we
can conclude $y_{s}\left(  k\right)  \rightarrow\mathcal{B(}\left\Vert
c\right\Vert \gamma(\delta)+\left\Vert c\right\Vert \nu)$ as $k\rightarrow
\infty.$ $\square$

\textbf{Remark 9}. Since the sensor sampling rate is often faster than the
control rate, $y$ is assumed to be measured continuously for the sake of
simplicity so that $\hat{x}_{p}\equiv x_{p}$ and $\hat{x}_{s}\equiv x_{s}$. In
(\ref{Pb3Observer}), the term $\int_{0}^{T_{s}}e^{As}\left(  y\left(
s\right)  -r\left(  s\right)  \right)  ds$ can be approximated more accurately
by using a finer sensor sampling rate.

\section{An Illustrative Example}

\subsection{Problem Formulation}

In this paper, a single-link robot arm with a revolute elastic joint rotating
in a vertical plane is served as an application \cite{Marino(1995)}:%
\begin{align}
\dot{x}  &  =A_{0}x+bu+\phi_{0}\left(  y\right)  +d,x\left(  0\right)
=x_{0}\nonumber\\
y  &  =c^{T}x. \label{RTACmodel1}%
\end{align}
Here%
\begin{equation}
A_{0}=\left[
\begin{array}
[c]{cccc}%
0 & 1 & 0 & 0\\
-\frac{K}{J_{l}} & -\frac{F_{l}}{J_{l}} & \frac{K}{J_{l}} & 0\\
0 & 0 & 0 & 1\\
\frac{K}{J_{m}} & 0 & -\frac{K}{J_{m}} & -\frac{F_{m}}{J_{m}}%
\end{array}
\right]  ,b=\left[
\begin{array}
[c]{c}%
0\\
0\\
0\\
1
\end{array}
\right]  ,c=\left[
\begin{array}
[c]{c}%
1\\
0\\
0\\
0
\end{array}
\right]  ,\phi_{0}\left(  y\right)  =\left[
\begin{array}
[c]{c}%
0\\
-\frac{Mgl}{J_{l}}\sin y\\
0\\
0
\end{array}
\right]  ,d=\left[
\begin{array}
[c]{c}%
0\\
d_{1}\\
0\\
d_{2}%
\end{array}
\right]  \label{par}%
\end{equation}
where $x=\left[
\begin{array}
[c]{cccc}%
x_{1} & x_{2} & x_{3} & x_{4}%
\end{array}
\right]  ^{T}$ are the link displacement (rad), link velocity (rad/s), rotor
displacement (rad) and rotor velocity (rad/s), respectively; $d_{1}$ and
$d_{2}$ are unknown disturbances. The initial value is assumed to be $x\left(
0\right)  =\left[
\begin{array}
[c]{cccc}%
0.05 & 0 & 0.05 & 0
\end{array}
\right]  ^{T}.$ Let the link inertia $J_{l}=2$ kg$\cdot$m$^{2},$ motor rotor
inertia $J_{m}=0.5$kg$\cdot$m$^{2},$ elastic constant $k=0.05$ kg$\cdot$%
m$^{2}$/s$,$ link mass $M=0.5$ kg, gravitational acceleration $g=9.8$
m/s$^{2}$, the center of mass $l=0.5$ m and viscous friction coefficients
$F_{l}=F_{m}=0.2$ kg$\cdot$m$^{2}$/s. The control $\tau$ is the torque
delivered by the motor. The control problem here is: assuming only $y$ is
measured, design $u$ so that $y$ tracks a smooth enough reference
$r\ $asymptotically or with a good tracking accuracy. It is easy to verify
that the pair $\left(  A_{0},c\right)  $ is observable. It is found that
$A_{0}$ is unstable. Choose $p=\left[
\begin{array}
[c]{cccc}%
-2.10 & -1.295 & -9.36 & 3.044
\end{array}
\right]  ^{T}.$ Then the system (\ref{RTACmodel1}) can be formulated as
(\ref{equ0}) with $A=A_{0}+pc^{T}\ $and$\ \phi\left(  y\right)  =\phi
_{0}\left(  y\right)  -py$, where $A$ is stable. Assume that the desired
trajectory is $r(t)=0.05\sin(\frac{2\pi}{T}t)+0.1$, while the periodic
disturbances are $d_{1}(t)=0.04\sin(\frac{2\pi}{T}t)$ and $d_{2}%
(t)=0.02\cos\left(  \frac{2\pi}{T}t\right)  \sin(\frac{2\pi}{T}t),$ where
$T=\frac{20\pi}{3}$s. Let the sampling period be $T_{s}=0.1$s. Then $N=209.$

\subsection{Controller design}

\subsubsection{Controller Design for Primary System}

Under the sampling period $T_{s}=0.1$s, the discrete-time transfer function
$P\left(  z\right)  $ in (\ref{discrete}) is%
\begin{equation}
P\left(  z\right)  =\frac{9.9\ast10^{-8}(z+9.399)(z+0.9493)(z+0.09589)}%
{(z-0.9512)(z-0.9418)(z-0.9324)(z-0.9231)} \label{P}%
\end{equation}
where there exists an unstable zero $-9.399.$ Therefore, $P\left(  z\right)  $
is non-minimum phase. So, the condition (\ref{condition0}) cannot be
satisfied. We further can obtain $T\left(  z\right)  $. According to $T\left(
z\right)  $ and the ZPETC algorithm (\ref{ZPETC_L}), we have $L\left(
z\right)  .$ Then%
\[
T\left(  z\right)  L\left(  z\right)  =z^{-1}\frac{(1+9.399z^{-1}%
)(1+9.399z)}{(1+9.399)^{2}}.
\]
Choose $Q\left(  z\right)  $ to be an FIR filter as%
\begin{equation}
Q\left(  z\right)  =0.5+0.2z^{-1}+0.2z^{-2}+0.1z^{-3}. \label{Q}%
\end{equation}
According to \cite{Steinbuch(2007)}, we choose%
\begin{equation}
W\left(  z\right)  =2-z^{-N} \label{W_design}%
\end{equation}
to improve robustness against the period variation. It is easy to check that
the closed-loop system is stable. Furthermore, to compare the robustness of
the tracking accuracy against the period variation, the amplitude of the
transfer function in (\ref{trackingerror}) with both $W\left(  z\right)  =1$
and $W\left(  z\right)  =2-z^{-N}$ are plotted in Fig. 2. As shown, although
the periodic components will be attenuated with $W\left(  z\right)  =1$ more
strongly, the higher-order RC with $W\left(  z\right)  =2-z^{-N}$ is less
sensitive to the period variation in the low frequency band. Since the low
frequency band is often dominant in the signal $r\left(  z\right)
-d_{r}\left(  z\right)  ,$ the higher-order RC can improve the robustness of
the tracking accuracy against the period variation. This will be confirmed
next.\begin{figure}[h]
\begin{center}
\includegraphics[
scale=0.65]{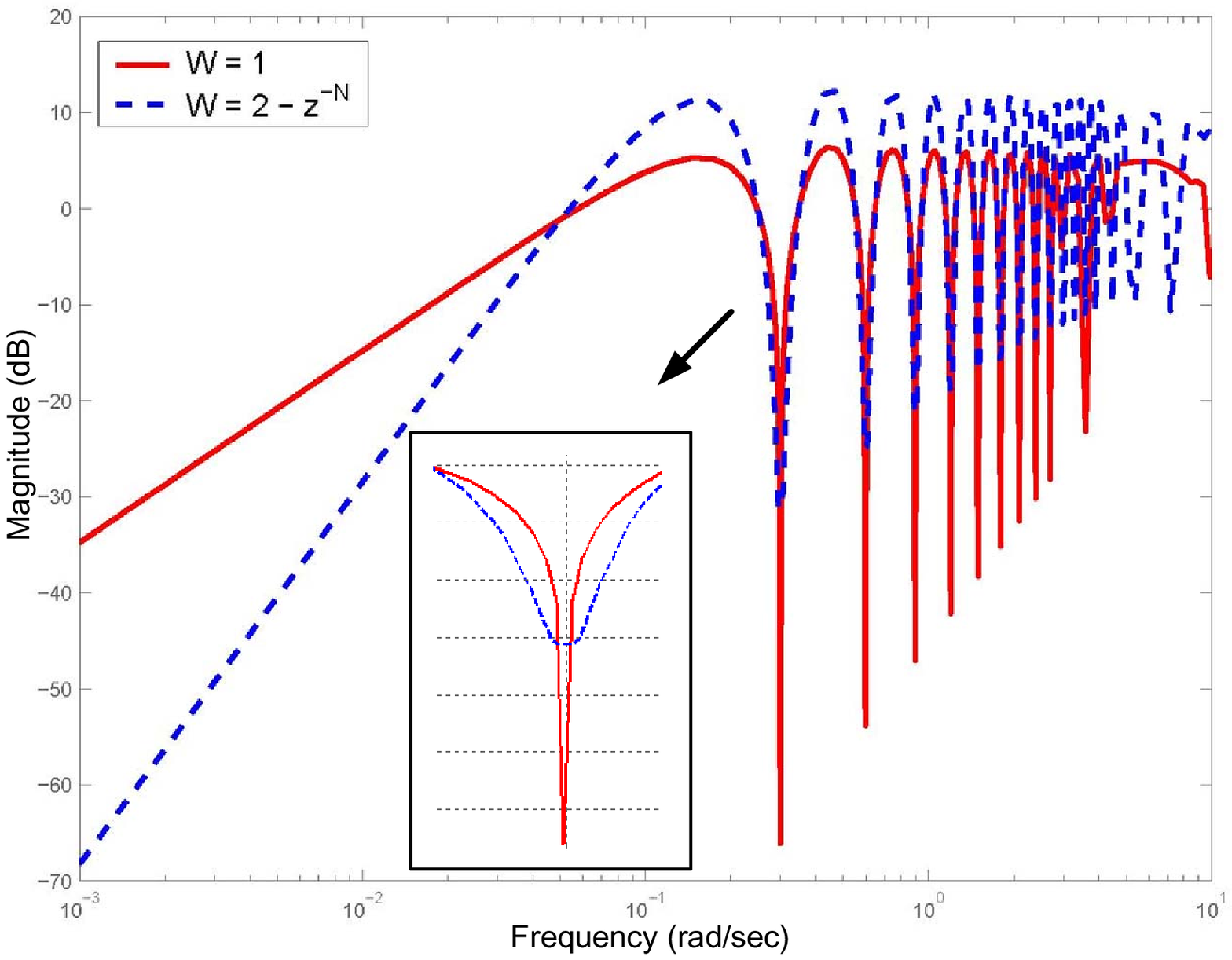}
\end{center}
\par
\vspace*{-18pt}\caption{The amplitude of $\frac{1}{1+P\left(  z\right)  }%
\frac{1-Q\left(  z\right)  W\left(  z\right)  z^{-N}}{1-Q\left(  z\right)
W\left(  z\right)  z^{-N}\left(  1-T\left(  z\right)  L\left(  z\right)
\right)  }$}%
\end{figure}\vspace*{-18pt}

\subsection{Controller Design for Secondary System}

For\textit{\ }the system (\ref{equ1_Sec}), by the backstepping technique
\cite{Khalil(2002)}, we design%
\begin{equation}
u_{s}\left(  x_{s}\right)  =\mu_{1}+\frac{J_{l}}{K}\left(  v+\mu_{2}\right)
\label{Assumption4_Sys_con}%
\end{equation}
where%
\begin{align*}
v  &  =-7.5x_{s,1}-19x_{s,2}-17\eta_{3}-7\eta_{4}\\
\mu_{1}  &  =-\eta_{3}+\frac{K}{J_{m}}x_{s,1}-\frac{K}{J_{m}}x_{s,3}%
-\frac{F_{m}}{J_{m}}x_{s,4}\\
\mu_{2}  &  =\frac{F_{l}}{J_{l}}\eta_{4}+\frac{Mgl}{J_{l}}\left(  \eta
_{3}+\ddot{r}\right)  \cos\left(  x_{s,1}+r\right)  -\frac{Mgl}{J_{l}}\left[
\left(  x_{s,2}+\dot{r}\right)  ^{2}\sin\left(  x_{s,1}+r\right)  +\ddot
{r}\cos\left(  r\right)  -\dot{r}^{2}\sin\left(  r\right)  \right] \\
\eta_{3}  &  =-\frac{F_{l}}{J_{l}}x_{s,2}-\frac{K}{J_{l}}\left(
x_{s,1}-x_{s,3}\right)  -\frac{Mgl}{J_{l}}\left[  \sin\left(  x_{s,1}%
+r\right)  -\sin\left(  r\right)  \right] \\
\eta_{4}  &  =-\frac{F_{l}}{J_{l}}\eta_{3}-\frac{K}{J_{l}}\left(
x_{s,2}-x_{s,4}\right)  -\frac{Mgl}{J_{l}}\left[  \left(  x_{s,2}+\dot
{r}\right)  \cos\left(  x_{s,1}+r\right)  -\dot{r}\cos\left(  r\right)
\right] \\
x_{s}  &  =\left[
\begin{array}
[c]{cccc}%
x_{s,1} & x_{s,2} & x_{s,3} & x_{s,4}%
\end{array}
\right]  ^{T}.
\end{align*}
The controller (\ref{Assumption4_Sys_con}) can solve \textit{Problem 2}. (The
design and proof are omitted for lake of space)\textit{.}

\subsection{Controller Integration and Simulation}

The final controller is given by\ (\ref{Pb3Controller}), where $L\left(
z\right)  ,Q\left(  z\right)  $ in $u_{p}$ are chosen as in \textit{Section
IV.B}, while $u_{s}$ is chosen as in (\ref{Assumption4_Sys_con}). The
variables $y_{p}$ and $x_{s}$ are estimated by the observer (\ref{Pb3Observer}%
) with the sensor sampling rate $T_{ss}=0.01s$. In the controller combination,
the variables $y_{p}$ and $x_{s}$ will be replaced with $\hat{y}_{p}$ and
$\hat{x}_{s}$. To compare the robustness of the tracking accuracy against the
period variation, both $W\left(  z\right)  =1$ and $W\left(  z\right)
=2-z^{-N}$ are taken into consideration, and the true period is assumed to be
$\frac{20\pi}{3}\left(  1+\alpha\right)  ,$ where $\alpha$ is the
perturbation. The tracking error is uniformly ultimately bounded. In Fig. 3,
the ultimate bound is plotted as a function of the perturbation $\alpha.$ As
shown, the ultimate bound is small if $\alpha$ is small. This implies that the
proposed discrete-time RC can drive $y$ to track $r.$ More importantly, the
ultimate bound of the steady-state tracking error produced by the proposed
higher-order RC with $W\left(  z\right)  =2-z^{-N}$ is less sensitive to the
perturbation $\alpha$ in comparison with that by the traditional RC with
$W\left(  z\right)  =1$. Therefore, we have achieved our initial goal that a
discrete-time output-feedback RC is designed for the nonlinear system
(\ref{equ0}) such that $y$ can track $r$ robustly with respect to the period
variation.\begin{figure}[h]
\begin{center}
\includegraphics[
scale=0.65]{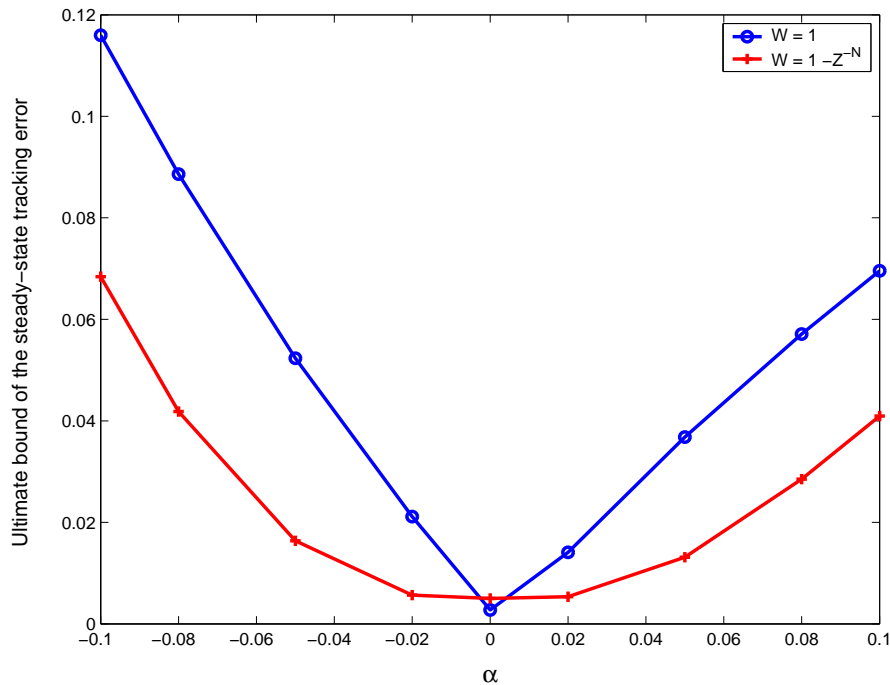}
\end{center}
\par
\vspace*{-18pt}\caption{The ultimate bound as function of the period-mismatch
with traditional RC and higher-order RC}%
\end{figure}

\section{Conclusions}

In this paper, a discrete-time output-feedback robust RC problem\ for\ a class
of systems with measurable nonlinearities was solved under the
additive-state-decomposition-based tracking control framework. To demonstrate
its effectiveness, the design method was applied to a single-link robot arm
with a revolute elastic joint rotating in a vertical plane. From the analysis
and simulation, the resulting discrete-time RC can track the periodic
reference and compensate for the periodic disturbance robustly with respect to
the period variation. While the higher-order RC and backstepping technique are
not new, our contribution is how to use additive state decomposition to solve
a new and challenging RC problem, namely the discrete-time output-feedback
robust RC problem for a class of nonlinear systems. Furthermore, we bridge RC
design in the frequency domain for LTI systems to include a class of nonlinear
systems; thus RC problems for certain nonlinear systems become tractable.

\end{document}